\documentclass[useAMS,usenatbib]{mn2e}
\usepackage{natbib}
 \usepackage{graphicx}
 \usepackage{color}
 \usepackage{amsmath}
 
 \newcommand\Gaia{\textit{Gaia}}

\newcommand{\unit}[1]{\ensuremath{\, \mathrm{#1}}}
 
\title[\textcolor{black}{An Analytical Model of Radiation-Induced Charge Transfer Inefficiency for CCD Detectors}]{\textcolor{black}{An Analytical Model of Radiation-Induced Charge Transfer Inefficiency for CCD Detectors}}
\author[A.~Short et al.]{A.~Short$^{1}$,\thanks{E-mail: ashort@esa.int
 }
 C.~Crowley$^{1}$,
 J.H.J.~de~Bruijne$^{1}$
 \textcolor{black}{T.~Prod'homme$^{1}$} 
 \\
$^{1}$European Space Agency, Keplerlaan 1, 2201 AG,  Noordwijk, The Netherlands\\ 
 }
\begin{document}

\date{in draft form 2012 Oct 30}

\pagerange{\pageref{firstpage}--\pageref{lastpage}} \pubyear{}

\maketitle

\label{firstpage}

\begin{abstract}
The European Space Agency's {\Gaia} mission is scheduled for launch in 2013. It will operate at L2 for 5 years, rotating slowly to scan the sky so that its two optical telescopes will repeatedly observe more than one billion stars. The resulting data set will be iteratively reduced to solve for the  position, parallax and proper motion of every observed star. The focal plane contains 106 large area silicon CCDs continuously operating in a mode where the line transfer rate and the satellite rotation are in synchronisation.

One of the greatest challenges facing the mission is radiation damage to the CCDs which will cause charge \textcolor{black}{deferral} and image shape distortion. This is particularly important because of the extreme accuracy requirements  of the mission. 
Despite steps taken at hardware level to minimise the effects of radiation,  the residual distortion will need to be calibrated during the pipeline data processing. Due to the volume and inhomogeneity of data involved, this requires a model which describes the effects of the radiation damage   which is physically realistic, yet fast enough to implement in the pipeline. The resulting charge distortion model was  developed specifically for the {\Gaia} CCD operating mode. However, a generalised  version is presented in this paper and this has already been applied  in a broader context, for example to investigate the  impact of radiation damage on  the Euclid dark-energy mission data.
\end{abstract}


\section{Introduction \label{sect:intro} }


{\Gaia} is a European Space Agency mission, planned for launch in  2013, which aims to perform global astrometry on approximately one percent of the estimated galactic stellar population to unprecedented accuracy\footnote{For science performance predictions see \cite{2012Ap&SS.tmp...68D}, for current updates on the science performances see http://www.rssd.esa.int/gaia}\citep[see][]{2001A&A...369..339P,2008IAUS..248..217L}.   The scanning satellite will operate at the earth/moon-sun Lagrangian point for 5 years, rotating slowly so that its two optical telescopes will repeatedly observe more than a thousand million stars. The resulting dataset will be iteratively reduced to solve for the position, parallax and proper motion of every observed star (additionally, dispersed spectra will be obtained for a  subset of these objects). The mission requirements for the astrometric precision are in the micro-arcsecond ($\umu$as) regime. These stringent astrometric requirements translate into milli-pixel-level accuracies on the stellar image location  per \textcolor{black}{ individual CCD observation, e.g., \citet{2011MNRAS.414.2215P}. }

The {\Gaia} focal plane \citep[e.g.,][]{2010SPIE.7731E..35D}, the largest to date to be sent to space,  contains 106 large-area silicon CCDs continuously operating in Time Delay Integration (TDI) mode at a line-clocking rate which is in synchronisation with the satellite rotation rate. Each CCD was custom-made by e2v technologies for the {\Gaia} project and consists of 4500 $\times$ 1966 (parallel $\times$ serial) pixels of physical dimension 10 $\times$ 30 $\umu \mathrm{m}$ \citep[for further detail see][]{2011MNRAS.414.2215P}. Due to a combination of the large number of transfers of the charge packet (parallel) and fast clocking speed (serial), the limits placed on the amount of radiation shielding due to mass-budget constraints, the predicted radiation environment at L2, and the extreme accuracy requirements on the image location estimates, \textcolor{black}{Charge Transfer Inefficiency (CTI) of the CCDs was identified as a challenge for the {\Gaia} mission at an early stage. }

CTI during the transfer of electron packets from pixel to pixel is due to the stochastic capture and release of signal electrons into and out of trapping sites in the silicon lattice structure. This will have two major effects upon {\Gaia} data:

\begin{enumerate}
\item the removal of charge from the signal within the windowed image\textcolor{black}{\footnote{In order to meet  telemetry constraints and  to minimise readout noise, windows are placed around each confirmed source and only data from within these windows are read out and sent to ground. The window dimensions depend on the magnitude of the source.} resulting in a loss of signal and a corresponding irretrievable degradation in the  signal-to-noise ratio.}


\item A distortion and centroid shift of the image due to the emission of electrons into charge packets different to those which they were captured from.
\end{enumerate}

\begin{figure}
\includegraphics[width=88mm]{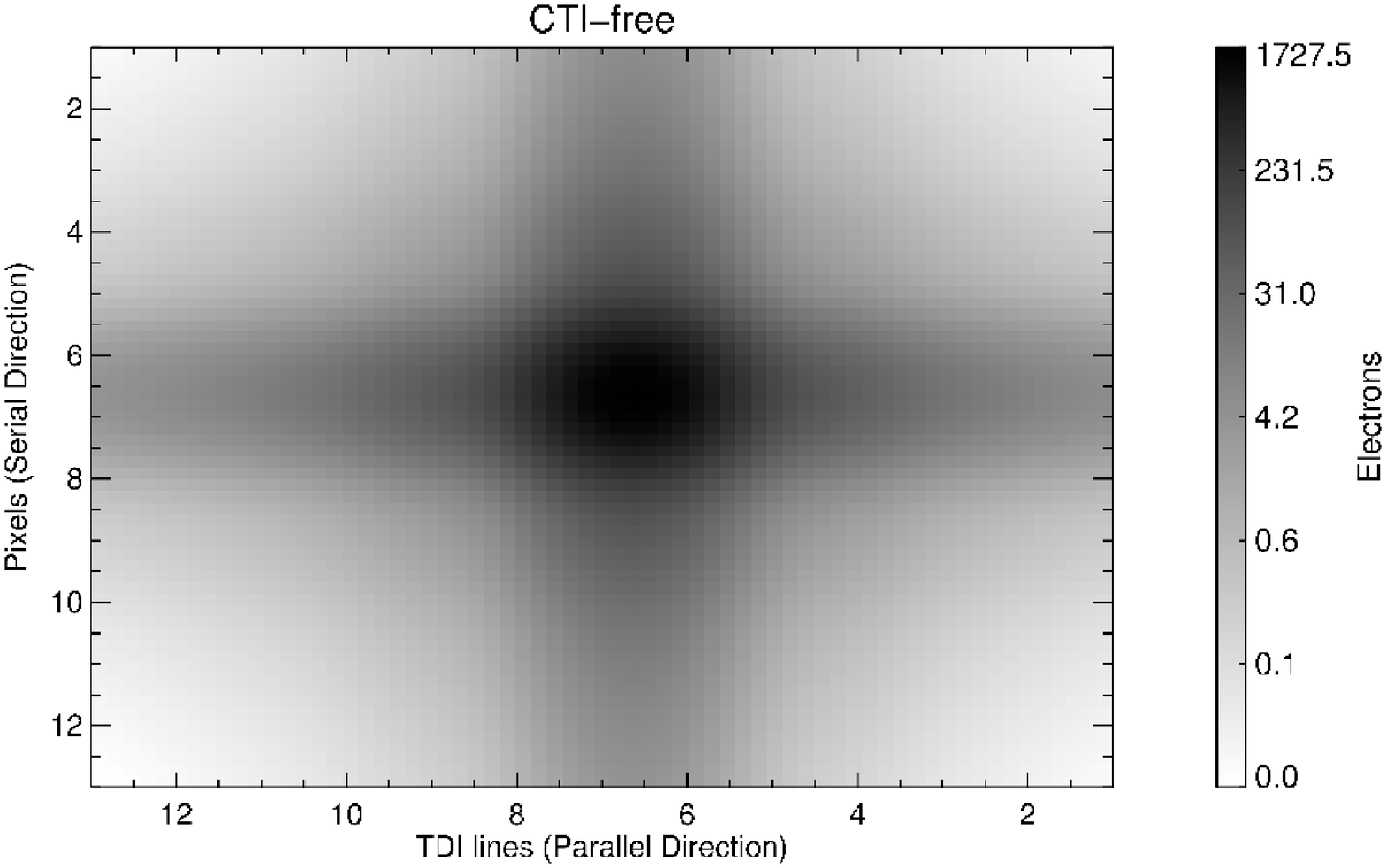}
\includegraphics[width=88mm]{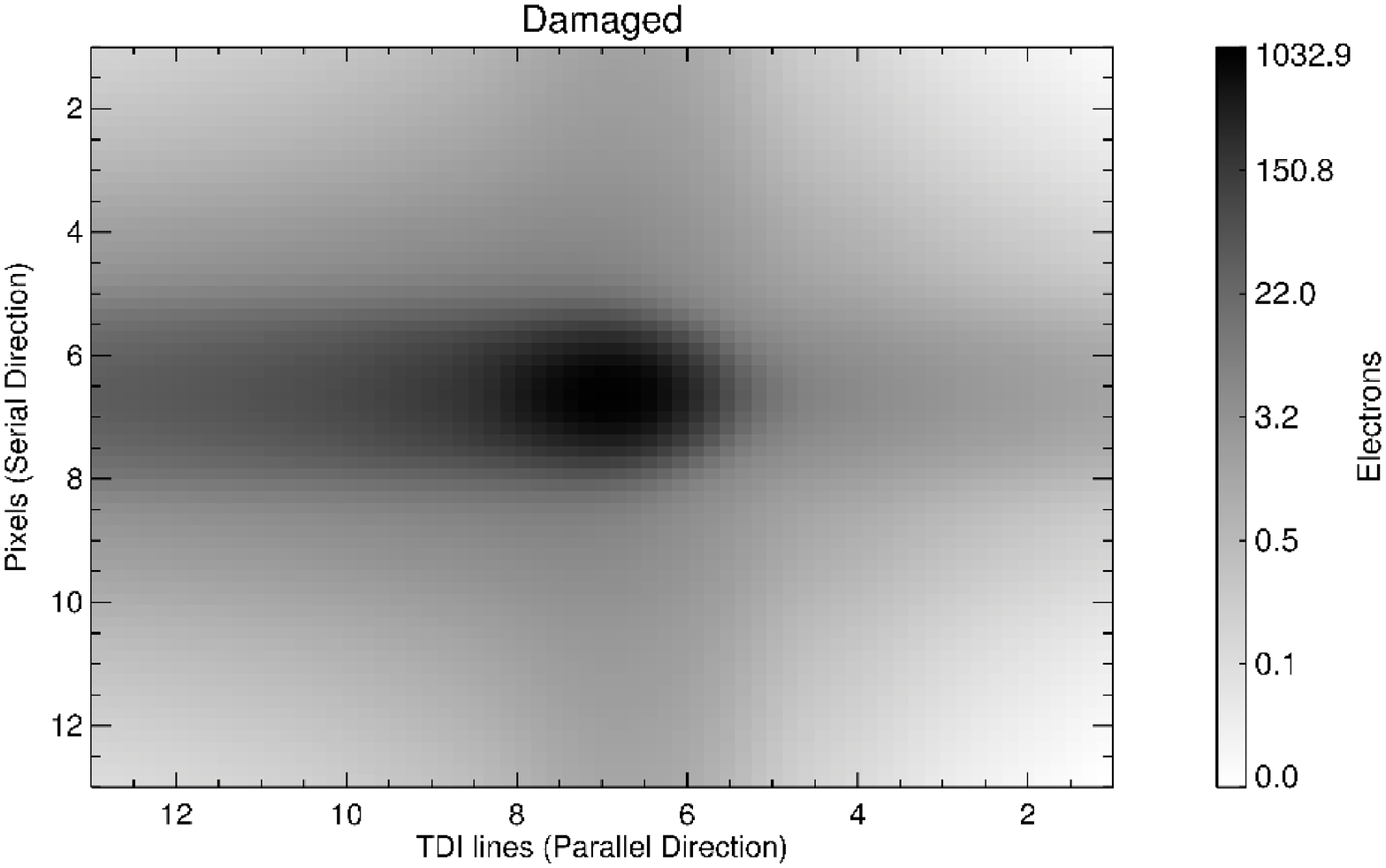}
\caption{ \textcolor{black}{{\bf{Upper Panel:}}  Simulation of a 5~$\times$~oversampled image in the abscence of CTI. {\bf{Lower Panel:}} The effect of parallel CTI on the image in the top panel. The CTI effect was simulated using the model described in this paper. Note that unrealistic model parameters were chosen in order to exaggerate the effect for clarity. The readout direction is to the right. } }
\label{fig:2ddamaged}
\end{figure}

\begin{figure}
\includegraphics[width=85mm]{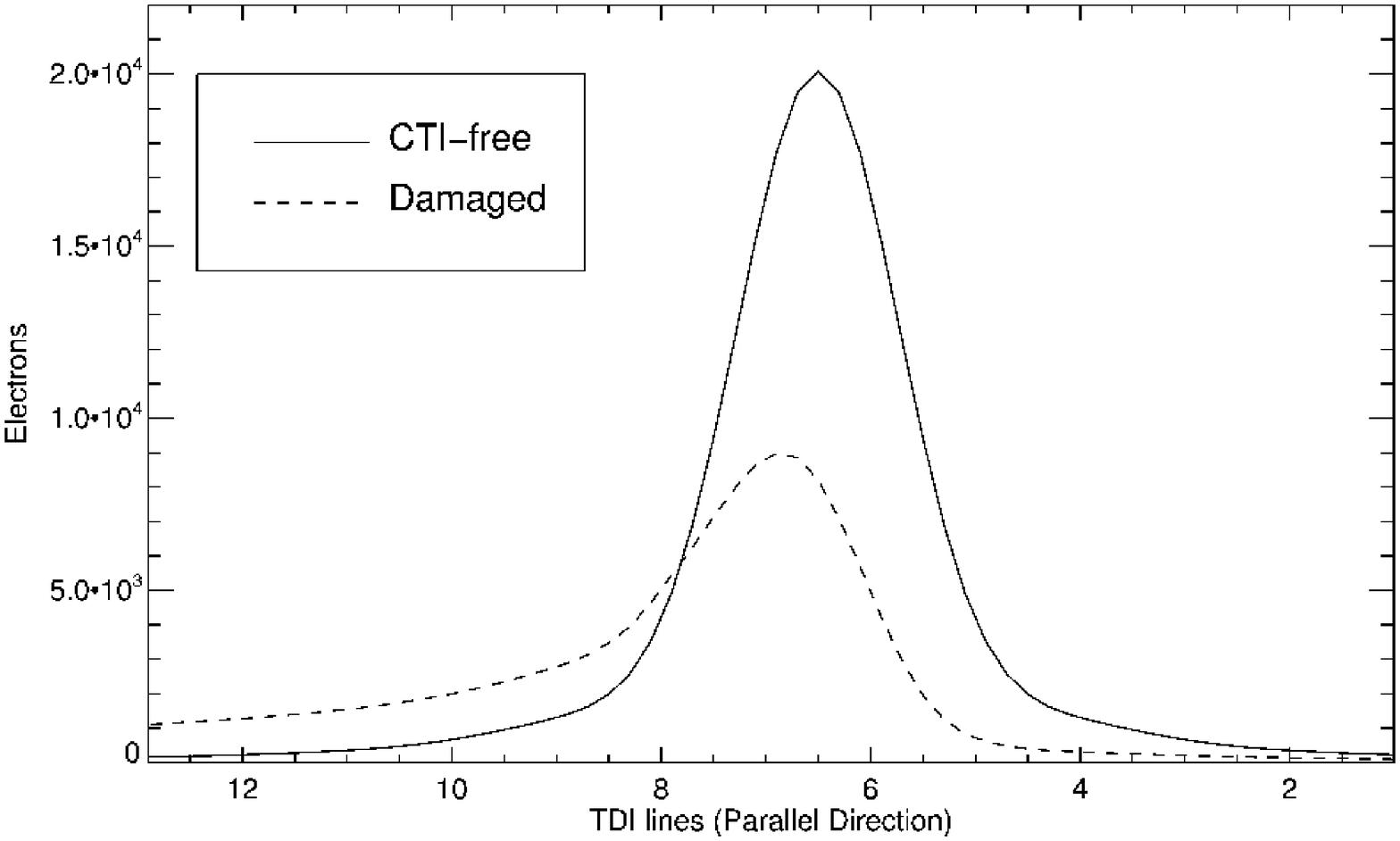}
\caption{Simulation of a 5~$\times$~oversampled image that is binned in the serial direction (the vast majority of {\Gaia} observations will have on-chip binning applied  in the serial direction). The effect of parallel CTI on the image is plotted in dashed. The CTI effect was simulated using the model described in this paper. Note that unrealistic  model parameters were chosen in order to exaggerate the effect for clarity. The readout direction is to the right.  }
\label{fig:1ddamaged}
\end{figure}

These effects are schematically illustrated in Figure~\ref{fig:2ddamaged} and Figure~\ref{fig:1ddamaged}.

The distortion of the image shape, if uncalibrated, will introduce a systematic bias on the measured image location and result in a degradation of the  end-of-mission astrometric accuracies from $\umu$as to mas levels. \textcolor{black}{Indeed, based on simulations, \citet{2011MNRAS.414.2215P} estimate a bias of up to  $\sim0.2$~pixels in the CCD location measurement ($\sim10$~mill-arcseconds) as a result of CTI in the uncalibrated case, which is in agreement with on-ground test data.} The main reasons that the effect on the performance is so large is due to the harsh radiation environment at L2 (dominated by energetic solar protons) and  the stringent mission  requirements on the image location measurements. It is well established that the interaction of high-energy protons with CCD devices can result in displacement damage in the silicon structure and the generation of `new' energy levels in the semiconductor band-gap, otherwise known as `traps', that can capture and, after some time, release the electrons.  These radiation-induced traps dramatically increase the CTI of the detector and raise the radiation issue to being   mission-critical in the absence of adequate mitigation. For further discussion of radiation-induced CTI see \citet{1998ITNS...45..154H,2001sccd.book.....J}.

In order to minimise the CTI effects, a number of hardware solutions has been implemented for {\Gaia}. For example, a mechanism for the periodic injection of charge into  lines of the CCD in order to keep traps filled has been implemented\footnote{\textcolor{black}{Note that these periodic charge injections will also act to decrease the scientific quality of the data in two main ways.  Some observations will be corrupted by the injection of charge across the acquired windowed data. Also, the trail from the injection will provide an additional background and thus inject additional noise into the data. However,  the benefits of the implementation of periodic charge injection outweigh these downsides (see the discussions in \citet{2011MNRAS.414.2215P} and \citet{ 2012MNRAS.422.2786H}). }} (charge injection). A supplementary buried channel has also been incorporated into the CCD design  to funnel charge from small electron packets into a smaller pixel volume during transfer in order to minimise the number of traps that will be encountered at low signals \citep[for further reading see][]{seabroke2012}. However, in order for {\Gaia} to meet its scientific requirements, a calibration of the residual effects of radiation-induced CTI is mandatory.

The entire set of {\Gaia}   observations will be used to  solve for  the astrometric parameters of each star, as well as the  satellite attitude parameters  and  calibration parameters  using the  Astrometric Global Iterative Solution (AGIS) described in \citet[][]{2012A&A...538A..78L}. Radiation-induced distortions of the images are a source of systematic error which, given a suitable model, could be removed as part of the global solution. Therefore,  the current baseline approach  to tackle the calibration of radiation-induced trapping effects in {\Gaia} data is through the application of a forward modelling technique in the pipeline processing.  Under this scheme, the parameters of the CTI-free image (location and integrated flux) and the parameters of a model which can replicate the CTI effects on the CTI-free image are iteratively adjusted until the predicted distorted image best matches the observed image. This approach requires a model that can replicate the distortion of a sampled image due to radiation-induced trapping effects, we call this a Charge Distortion Model (CDM). The forward-modelling approach is preferable to a corrective approach in order to preserve the noise properties of the data, as well as having the flexibility  to treat dispersed spectra and non-point source objects in the same manner as  isolated single stars in the data processing. However, the CDM must be realistic and flexible enough to reproduce the distortion effectively, whilst simultaneously being fast enough for implementation in the data processing pipeline. The CDM can be empirical and need not be based upon detailed physical modelling of electron trapping and release. However, any model based upon physical considerations is certain to be superior in terms of application over a broad parameter space using the fewest possible variables. Since computational speed is of high concern, the model need not treat individual pixel transfers, but instead, approximate the trapping and release during the transit of an entire CCD column (or, indeed, the readout register) in a single calculation. 
In order to treat the non-integrating (non-TDI) charge injection lines, transfer through the serial register and also for general applicability, the model should also have the capability to treat the charge transfer when the signal is non-integrating (imaging mode). 

During the course of {\Gaia} development and preparation studies, a great deal has been learned regarding the phenomenological interpretation of trapping, de-trapping and electron confinement from the very large to the very small signal regimes. Over the course of five radiation test campaigns, many datasets have been acquired in \textcolor{black}{TDI mode using partially-irradiated {\Gaia} CCDs} in order to characterise the CTI effects. Monte-Carlo \citep[e.g.,][]{2011MNRAS.414.2215P} and analytical models have also been developed which go some way to reproducing {\Gaia} radiation test data. \textcolor{black}{However, the sub-pixel level Monte Carlo CCD models developed for {\Gaia} are far too computationally expensive to be applied as a CDM in the forward modelling scheme which needs to run on $\sim10^{12}$ separate CCD observations collected over the course of the mission\footnote{\textcolor{black}{In order to demonstrate the speed advantages of this model,  a typical {\Gaia} CCD observation ($6 \times 12$ pixel image) was run through a Java implementation of this CDM (not optimised for speed). With 7 trap species used by the model the  it takes $\sim$~0.25 milli-seconds to process using a standard laptop with a Intel Core i5 processing chip. Running the same image through a Monte Carlo model \citep{2011MNRAS.414.2215P} takes $\sim$~60 seconds.} }} 
 In this paper, we present a model, based upon a physical understanding of trapping effects, but with some very significant simplifications such that it can be incorporated into the  data processing pipeline. \textcolor{black}{Further discussions and quantitative predictions on the performance of {\Gaia} in the presence of CTI (including the use of this CDM for CTI-mitigation) can be found within \citet{2011MNRAS.414.2215P} and \citet{ 2012MNRAS.422.2786H}.} 

The paper is structured as follows. In  Section~\ref{sect:modelbackground} we address the constraints imposed on the model from the analysis of {\Gaia}-specific test data acquired in TDI mode. The details of the model are then presented in Section~\ref{sect:model}. In Section~\ref{sect:data} we show an overview of the application of the model to test data and, finally, in Section~\ref{sect:conclusions} we present our conclusions.

\section{Background to the Model} \label{sect:modelbackground}

The objective is to derive an analytical CDM  that takes into account recent illumination and charge injection history\footnote{\textcolor{black}{The number of  electron-trapping sites that can capture an electron will of course depend upon the fraction of these sites that are currently vacant, the trap occupancy. The trap occupancy will thus depend on factors such as the time since last charge injection, the constant diffuse optical background incident upon the detector, the recent transit history of other light sources across the pixels in question etc.}} and is based upon accepted physical theory, i.e. Shockley-Read-Hall theory \citep[e.g.,][]{1998ITNS...45..154H} and knowledge gained from {\Gaia} testing. The first iteration of such a  CDM  was proposed by \cite{LL:ll-075}, however, his model is not based on physical theory and was always intended as a place-holder.

In order to apply the  Shockley-Read-Hall equations it is necessary to model the volume and density of the electron packets as they move through the CCD. This will give the numbers of traps encountered by an electron packet according to the trap density, as well as the probability that a trap captures or releases an electron. In this respect, the models developed for {\Gaia} fall into two broad categories: \emph{Confinement Volume} (or volume driven) models and \emph{Density Distribution} (or density driven) models.

CCD models usually consider electrons to be contained in charge packets resting in potential minima defined by the applied electrode potentials. The electron cloud is typically assumed to have a finite confinement volume which increases as more electrons are added. Hence a model such as:

\begin{align}
\frac{V_c}{V_g} &= \left(\frac{N_e}{FWC}\right)^\beta
\label{conf_1}
\end{align}

might be adopted for the volume of the charge cloud ($V_c$), where $N_e$ is the number of electrons in a pixel, $FWC$ is the pixel $Full$~$Well$~$Capacity$ in electrons and $V_g$ is the assumed maximum geometrical volume that electrons can occupy within a pixel (i.e. the volume of the electron cloud when $N_e=FWC$)\footnote{\textcolor{black}{A different function describing the behaviour of the confinement volume with electron packet size could also be used  which could be optimised for a specific CCD pixel architecture, however here we present a generalised model.  In fact, this function could possibly be optimised for use in the {\Gaia} data processing after analysis of real CTI-affected {\Gaia}  data at some stage during the mission.}}.  In most models, the value of $\beta$ would be assumed to be close to 1, the density of confined electrons would be almost constant and high enough for trapping within the electron packet volume to be considered instantaneous. We may refer to models with these properties  as volume driven models where all vacant traps within the electron packet will capture an electron whilst those outside cannot, and the amount of trapping observed will be driven by the electron packet volume. Early models developed for {\Gaia} were all volume driven. However, \textcolor{black}{there was a major shortcoming of this assumption which causes them to fail for {\Gaia}}.



Volume driven models are unable to reproduce or to explain the observed effect that a small number of background electrons can have upon overall trap occupancy. It is observed that very low level optical background (fat zero) is able to suppress a disproportionate fraction of slow traps \citep[e.g.,][]{LL:SWB-001, LL:AS-011}. According to volume driven models, this would require that background electrons must occupy a volume many thousands of times larger than an equivalent number of signal electrons. In addition, {\Gaia} testing clearly indicates that a diffuse optical background preferentially eliminates traps with long release time constants (slow traps) but does little or nothing to suppress faster traps. The effect of background upon the traps suggests that a different model must be applied allowing electrons to reach the traps in the first place. In other words, we must assume that $\beta$ in equation \ref{conf_1} is closer to 0 rather than 1 and that it is primarily the density of electrons rather than the volume which increases as more electrons are added. In this way, all electron packets will encounter a similar population of traps but the probability that a given trap captures an electron in a given time will depend upon the local electron density which increases with signal size.

\section{Model Description} \label{sect:model}

In this section we break down the model description into four differerent sections. We describe the modelling of the electron trapping in TDI mode, the trapping in imaging mode,  electon release and, finally,  the tracking of the trap occupancies as a function of time\footnote{\textcolor{black}{It should be pointed out that since the model is analytical,  any  trap parameters derived with this model will not necessarily correspond to exact physical quantities, even when the model can acurately reproduce what is observed in the data. They should therefore be thought of as effective, rather than physical, parameters.}}.

\subsection{Electron trapping in TDI mode} \label{sect:tdi}

\subsubsection{Interaction volume}

\begin{figure}
\includegraphics[width=85mm]{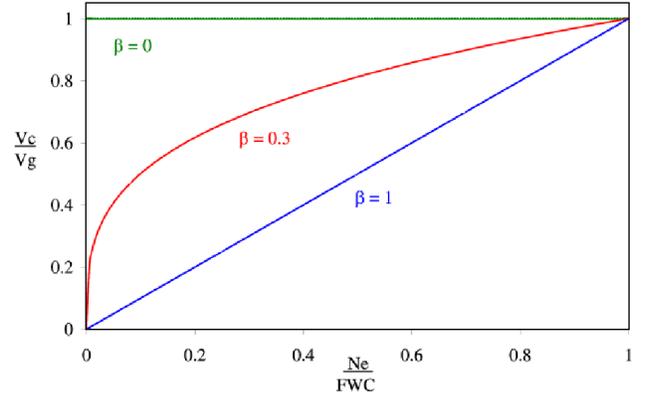}
\caption{Variation of electron confinement volume, $V_c$ with number of electrons, $N_e$ according to equation \ref{conf_1}.}
\label{fig:Vc}
\end{figure}

We take equation \ref{conf_1} as a starting point, noting that $\beta$ is not constrained to values close to 1 but may take any value between 0 and 1 allowing complete flexibility with regard to the question of volume or density driven models. Clearly, if $\beta$ is equal to 0 then $V_c=V_g$ is a constant. Only the electron density increases as electrons are added and trapping is entirely density driven. If $\beta$ is equal to 1 then $V_c$ grows linearly with $N_e$, electron density is constant and trapping is entirely driven by the interaction volume of the electrons with the traps. Figure \ref{fig:Vc} illustrates the variation of electron confinement volume, $V_c$ as a function of $N_e$ and $\beta$ (according to equation \ref{conf_1}).

Since we intend to treat the trapping along the column in a single step, we may derive the fraction of the total volume of silicon in the column that was ``seen'' by a given sample ($S$), by integrating equation \ref{conf_1} as the signal electrons accumulate along the CCD column\footnote{\textcolor{black}{By `column' we mean the column of pixels in the parallel direction that an electron packet must be transferred  along before reaching the readout register (also called the serial register).}} from $N_e=0$ to $N_e=S$:

\begin{align}
\frac{\mbox{volume that sees electrons}}{\mbox{volume of column}} &= \frac{1}{S}\int_0^{S}  \left(\frac{N_e}{FWC}\right)^\beta  \,\mathrm{d}N_e  \nonumber \\ 
%
 &= \frac{1}{1+\beta}  \left(\frac{S}{FWC}\right)^\beta
\label{ifrac2}
\end{align}

where it should be understood that $S$, the input to the CDM is the size of the sample (in electrons) that would be read out of the CCD in the absence of trapping, i.e. the un-distorted sample. Assuming that $N_T$ traps are distributed uniformly throughout the volume of the column, then the fraction of these traps with which the $S$ electrons can interact is equal to the interacting fraction of the column volume. However, if we assume that $N_o$ of the traps within the column are already occupied, then we can write the fraction of all traps in the column that will interact with the $S$ electrons {\it{and}} that are vacant\footnote{\textcolor{black}{It should be noted that $N_o$ describes the state of the traps in a column as seen by a given TDI sample and so refers to the state of each trap along the column at the moment in time when it is encountered by the current sample, i.e., immediately after being left by the previous sample. Therefore, it is not referring to the state of all of the traps in the column at the moment when a new sample begins to transit the CCD.}}:

\begin{align}
F_T &= \frac{\mbox{vacant traps that see electrons}}{\mbox{total number of traps}}  \nonumber \\
&= \frac{1}{1+\beta}  \left(\frac{S}{FWC}\right)^\beta - \frac{N_o}{N_T} 
\label{ifrac3}
\end{align}

Similarly, we may write the fraction of signal electrons, $S$, that see vacant (rather than occupied) traps:

\begin{align}
F_e &= \frac{\mbox{electrons that see vacant traps}}{\mbox{total number of electrons}} \nonumber \\
&= \frac{\frac{1}{1+\beta}  \left(\frac{S}{FWC}\right)^\beta - \frac{N_o}{N_T}}{\frac{1}{1+\beta}  \left(\frac{S}{FWC}\right)^\beta}
\label{ifrac4}
\end{align}

If trapping were allowed to proceed until all of the vacant traps were occupied or all of the available electrons were captured, then the number of captures, $N_c$ would be equal to the number of available traps or the number of available electrons (whichever is fewer). This may be expressed:

\begin{align*}
N_c|_{t\rightarrow \infty} &= \frac{N_TF_TSF_e}{N_TF_T+SF_e} \nonumber \\
&= \frac{N_T\left(\frac{1}{1+\beta}  \left(\frac{S}{FWC}\right)^\beta - \frac{N_o}{N_T}\right)S\frac{\left(\frac{1}{1+\beta}  \left(\frac{S}{FWC}\right)^\beta - \frac{N_o}{N_T}\right)}{\frac{1}{1+\beta}  \left(\frac{S}{FWC}\right)^\beta}}{N_T\left(\frac{1}{1+\beta}  \left(\frac{S}{FWC}\right)^\beta - \frac{N_o}{N_T}\right)+S\frac{\left(\frac{1}{1+\beta}  \left(\frac{S}{FWC}\right)^\beta - \frac{N_o}{N_T}\right)}{\frac{1}{1+\beta}  \left(\frac{S}{FWC}\right)^\beta}} \nonumber \\
\end{align*}

%

\begin{align}
=\frac{\gamma  S^\beta - N_o}{\gamma S^{\beta-1}+1}\qquad \mbox{where} \qquad \gamma=\frac{N_T}{\left(1+\beta\right)FWC^\beta}
\label{ifrac8}
\end{align}

When implementing the model, it is more general to use a trap density, $n_t$, rather than a number of traps in the column, $N_T$. Thus, substituting $N_T=2n_tV_gx$ where $x$ is the number of TDI transfers or the column length in pixels, we get:

\begin{equation}
\gamma=\frac{2n_tV_gx}{\left(1+\beta\right)FWC^\beta}
\label{ifrac9}
\end{equation}


\subsubsection{Finite trapping time}

Equation \ref{ifrac8} may be used to model electron trapping in TDI mode in the event that capture times are always short in relation to electron-trap interaction times (i.e. semi instantaneous trapping). However, it is well established through {\Gaia} testing, that the very low electron densities being transferred in TDI mode give rise to electron capture times that are on the order of the pixel dwell time and it is therefore not appropriate to consider instantaneous trapping. Instead, we must evaluate the probability of electron capture according to electron density. In TDI mode the electron density is increasing as the signal integrates along the CCD column. Therefore, most {\Gaia} models evaluate the probability of capture (and release) per line transfer or intra-pixel CCD phase (4\,500 transfers or 18\,000 phases, since each pixel contains four individual gate electrodes). However, for the purposes of this high speed model, we require an expression giving the column-averaged probability of capture for a transit as a function of (integrating) signal size and trap occupancy.

If we assume that traps in the CCD column interact with the electrons of a given sample for a period equal to half the TDI period ($t$) then the probability ($P_c$) that a vacant trap will capture an electron is:

\begin{equation}
P_c=1-\mbox{e}^{-\frac{t}{2\tau_c}}
\label{prob}
\end{equation}

where $\tau_c$ is the capture time constant:

\begin{equation}
\tau_c=\frac{1}{\sigma v_{t} n_e}
\label{capture_const}
\end{equation}

The quantity $\sigma$ is the trap capture cross-section, $n_e$ is the electron density in the vicinity of the trap and $v_{t}$ is the electron thermal velocity:

\begin{equation}
\frac{1}{2}m_e^*v_{t}^2=\frac{3}{2}kT \qquad \Rightarrow \qquad v_{t}=\sqrt{\frac{3kT}{m_e^*}}
\label{thermal_v}
\end{equation}

where $m_e^*$ is the \emph{effective} electron mass in silicon which approximately equals half the free electron rest mass\footnote{\textcolor{black}{Particles in a specific solid material, and under the influence of an external electromagnetic field, can generally be approximated to behave as if they were free particles  under semi-classical solid-state physics, however, a modified mass needs to be used, which is the effective mass for that specific material.}}. 

If we assume that a number of electrons ($N_e$) are contained within an effective confinement volume ($V_c$), then:

\begin{equation}
\tau_c=\frac{V_c}{\sigma v_{t} N_e}
\label{capture_const2}
\end{equation}

Substituting into equation \ref{prob}:

\begin{equation}
P_c=1- \exp\left( -\frac{t\sigma v_{t} N_e}{2V_c} \right)
\label{prob2}
\end{equation}

and substituting for $V_c$ from equation \ref{conf_1}:

\begin{equation}
P_c=1- \exp\left( -\frac{t\sigma v_{t} N_e}{2V_g}  \left(\frac{FWC}{N_e}\right)^\beta  \right)
\label{prob3}
\end{equation}

Assuming that the TDI period, $t$, is constant, we then find:

\begin{equation}
P_c(N_e)=1-\mbox{e}^{-\alpha N_e^{1-\beta}} \qquad \mbox{where} \qquad \alpha=\frac{t\sigma v_{t}FWC^\beta}{2V_g}
\label{prob4}
\end{equation}

Equation \ref{prob4} gives the capture probability (per vacant trap) as a function of the number of sample electrons, $N_e$. However, we require an effective capture probability for the entire column in which the number of electrons is increasing from $N_e=0$ to $N_e=S$ where $S$  is the size of the sample in the absence of trapping. To first order, we may evaluate an effective or column-average capture probability as a function of the input signal, $S$, using:

\begin{align}
\mbox{column average $P_c(N_e)$} &=\bar{P_c}(S) \nonumber \\
&= \frac{1}{S}\int_0^{S} P_c(N_e)   \,\mathrm{d}N_e  \nonumber \\
&=\frac{1}{S}\int_0^{S}  1-\mbox{e}^{-\alpha N_e^{1-\beta}}  \,\mathrm{d}N_e
\label{int1a}
\end{align}


which yields:

\begin{align}
\bar{P_c}(S) &= \nonumber 
\end{align}

\begin{align}
1-\frac{\alpha^{\frac{1}{\beta-1}}\Gamma\left(1+\frac{1}{1-\beta}\right)}{S}-\frac{\left(\alpha S^{1-\beta}\right)^{\left(\frac{\beta}{\beta-1}-1\right)}\Gamma\left(1-\frac{\beta}{\beta-1},\alpha S^{1-\beta}\right)}{\beta-1}
\label{P_c}
\end{align}

where

\begin{equation}
\Gamma\left(z\right)=\int_0^\infty u^{z-1}\mbox{e}^{-u} \,\mathrm{d}u
\label{gamma}
\end{equation}

is the gamma function and

\begin{equation}
\Gamma\left(z,\xi\right)=\int_\xi^\infty u^{z-1}\mbox{e}^{-u} \,\mathrm{d}u
\label{gamma2}
\end{equation}

is the incomplete gamma function. Equation \ref{P_c} may be used in a model but it is unwieldy \textcolor{black}{and slow computationally} to implement, especially when attempting to evaluate the initial (or equilibrium) trap occupancy prior to the transit of the first signal packet. Instead, we make a very simple approximation that $\bar{P_c}(S) \approx P_c(S/2)$ and then test the validity by comparison with equation \ref{P_c}. Thus

\begin{equation}
\mbox{assume }\bar{P_c}(S) \approx 1-\exp\left\{-\alpha \left(\frac{S}{2}\right)^{1-\beta}\right\}.
\label{P_c2}
\end{equation}

\begin{figure}
\includegraphics[width=85mm]{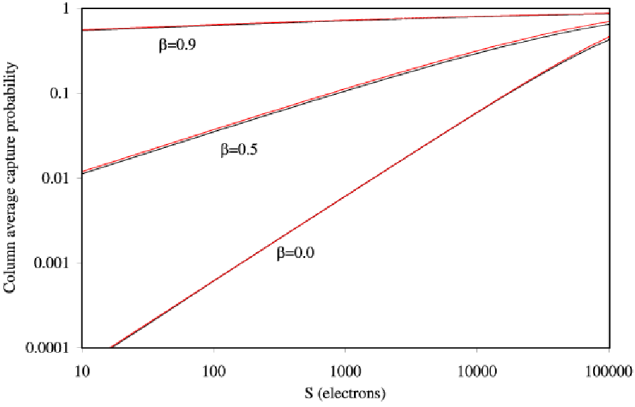}
\caption{Column-average electron capture probability, $\bar{P_c}(S)$, according to equation \ref{P_c} (black) and the simple approximation given in equation \ref{P_c2} (red). Note that a rather small value of capture cross-section ($3\times 10^{-18}$ cm$^2$) has been used in this example to illustrate the dependence over a large signal range. \textcolor{black}{Note that the difference is smaller for larger values of the capture cross section where the curve would be almost flat and close to unity.}}
\label{figure2}
\end{figure}

Figure \ref{figure2} illustrates equation \ref{P_c} (black) and the simple approximation given in equation \ref{P_c2} (red). The agreement is certainly good enough to justify the use of the simpler expression. Hence, we may combine equations \ref{ifrac8} and \ref{P_c2} to give an expression for the number of electrons, $N_c$, that will be captured by a given trap species during the transit of an integrating signal packet, $S$, as a function of physical trap parameters and the initial trap occupancy, $N_o$:

\begin{equation}
N_c=  \frac{\gamma  S^\beta - N_o}{\gamma S^{\beta-1}+1} \left(1-\exp\left\{-\alpha \left(\frac{S}{2}\right)^{1-\beta}\right\}\right) 
\label{N_c}
\end{equation}

Note that equation \ref{N_c} can give negative trapping when the number of occupied traps exceeds the number of interacting electrons and vacant traps. If the de-trapping process were the reverse of the trapping process, then this could be used directly to model trapping and de-trapping. However, trapping and de-trapping time constants are very different so this is not the case. When implementing equation \ref{N_c}, it is therefore necessary to avoid negative trapping. Thus, in summary, we have a model for electron capture by one trap species in TDI mode:

\begin{equation}
N_c=  \mbox{MAX}\left[\frac{\gamma  S^\beta - N_o}{\gamma S^{\beta-1}+1} \left(1-\exp\left\{-\alpha \left(\frac{S}{2}\right)^{1-\beta}\right\}\right)\:,\:0\right]
\label{N_c3}
\end{equation}

\begin{equation}
\mbox{where} \qquad \gamma=\frac{2n_tV_gx}{\left(1+\beta\right)FWC^\beta} \qquad \mbox{and} \qquad \alpha=\frac{t\sigma v_{t}FWC^\beta}{2V_g}
\label{N_c2}
\end{equation}

\subsection{Electron trapping in imaging mode (non-TDI)} \label{sect:imaging}

There are several circumstances where it is necessary to model trapping in non-TDI mode. These include the treatment of charge-injection lines which are injected at the top of the CCD and do not integrate along the CCD, modelling transfer along the serial register and of course applying the model to other missions which are not operating in TDI mode at all.

There are only a couple of differences in the final equations related to the fact that the signal is not integrating during the transfer. In imaging mode, the model may be summarised:

\begin{equation}
N_c=  \mbox{MAX}\left[\frac{\gamma  S^\beta - N_o}{\gamma S^{\beta-1}+1} \left(1-\exp\left\{-\alpha S^{1-\beta}\right\}\right)\:,\:0\right]
\label{N_c_i_1}
\end{equation}

\begin{equation}
\mbox{where} \qquad \gamma=\frac{2n_tV_g}{FWC^\beta}\left(x+i\right) \qquad \mbox{and} \qquad \alpha=\frac{t\sigma v_{t}FWC^\beta}{2V_g}
\label{N_c_i_2}
\end{equation}

where $x$ indicates the position on the CCD (in pixels) of the image to be processed, i.e., small $x$ for an image close to the readout node and large $x$ for objects far from the readout node and $i$ is the pixel coordinate of each sample within the image to be processed.

\subsection{Electron release} \label{sect:release}

If we assume that every occupied trap in the column has a time equal to one TDI period ($t$) in which to release the captured electron into a given sample, then the probability, $P_r$, that the trap will release the electron into the sample is simply:

\begin{equation}
P_r=1-\mbox{e}^{-\frac{t}{\tau_r}}
\label{P_r}
\end{equation}

where $\tau_r$ is the trap release time constant. The number of electrons released into the sample during a transit along the column is then simply:

\begin{equation}
N_r=N_o\left\{1-\mbox{e}^{-\frac{t}{\tau_r}}\right\}
\label{N_r}
\end{equation}

\subsection{Column state parameter and initial state} \label{sect:initial}


In implementing any CCD trapping model, it is very important to consider the state or occupancy of the traps (the value of $N_o$) at the beginning of the simulation. In the event that a continuous, unbroken stream of realistic data (including background etc.) were to be simulated (much longer than the capture or release time constants of the slowest traps), it would be sufficient to start with all traps empty or all traps full and to disregard the first few minutes of simulated data rather than to explicitly initialize $N_o$ with a realistic value. However, this is rarely practical, the state of the slower traps prior to the transit of a source will be dominated by two main factors. These are the time since the last charge injection  and the level of constant or slowly varying optical background (some fraction of slow traps kept full). We may therefore treat these two dependencies explicitly in order to quickly determine the state of traps prior to the transit of an astronomical source (or test data image).

\subsubsection{Diffuse optical background}

The effect of a constant (or slowly varying) low-level optical background upon the occupancy of slow traps has been demonstrated  during the {\Gaia} radiation test campaigns \textcolor{black}{(see Section~\ref{sect:modelbackground})}.  This reveals information about the temporal dependency or dynamics of trapping since the exposure time of traps to a constant background is radically different from the exposure time to transient sources. It also reveals something about the confinement volume of electrons within pixels since low levels of background appear capable of filling entire trap species.

Considering TDI mode for {\Gaia}, the initial trap occupancy, $N_o|_{init}$ due to a constant optical background is determined as follows. It is assumed that in the continuous presence of background electrons, traps will reach an equilibrium state of occupancy when $N_c=N_r$. Thus:

\begin{align}
\frac{\gamma  S_{dob}^\beta - N_o|_{init}}{\gamma S_{dob}^{\beta-1}+1} \left(1-\exp\left\{-\alpha \left(\frac{S_{dob}}{2}\right)^{1-\beta}\right\}\right) \nonumber \\
= N_o|_{init}\left\{1-\mbox{e}^{-\frac{t}{\tau_r}}\right\}
\label{equil}
\end{align}

where $S_{dob}$ is the diffuse optical background level in electrons per pixel measured at the readout register.

Solving equation \ref{equil} for $N_o|_{init}$ gives:

\begin{equation}
N_o|_{init}=\frac{AB}{A+C}
\label{N_o_init}
\end{equation}

where:

\begin{align}
\hspace{2cm} & A =\frac{1-\exp\left\{-\alpha \left(\frac{S_{dob}}{2}\right)^{1-\beta}\right\}}{\gamma S_{dob}^{\beta-1}+1}\qquad \qquad \text{and} \nonumber \\
\hfill & B =\gamma  S_{dob}^\beta \qquad \qquad \text{and} \nonumber \\
\hfill & C =1-\mbox{e}^{-\frac{t}{\tau_r}} 
\label{AB}
\end{align}

\subsubsection{Time since charge injection}

In the case of a high level of charge injection it may be sufficient to assume that the injection fills all traps which then subsequently empty at a rate determined by the exponential trap release time constant until they reach the occupancy level determined by the diffuse optical background. For example, adding an exponential term to equation \ref{N_o_init}, the initial state of the column would be given by:

\begin{equation}
N_o|_{init}= 2n_tV_gx\mbox{e}^{-\frac{t_{ci}}{\tau_r}} +  \frac{AB}{A+C}\left(1-\mbox{e}^{-\frac{t_{ci}}{\tau_r}} \right)
\label{N_o_init2}
\end{equation}

where $t_{ci}$ is the time elapsed since charge injection and $\tau_r$ is the trap release time constant.

However, in the case of lower levels of charge injection where it may not be adequate to assume that all traps are saturated, it is better to simulate the capture of electrons from the charge injection lines explicitly to determine the fraction of traps that are filled. These then empty according to the exponential trap release time constant as before.  In order to simulate charge injection correctly, it is necessary to consider that charge injection is effectively in image mode rather than TDI mode.

\section{Examples of Modelling Results} \label{sect:data}

\begin{figure*}
\includegraphics[width=80mm]{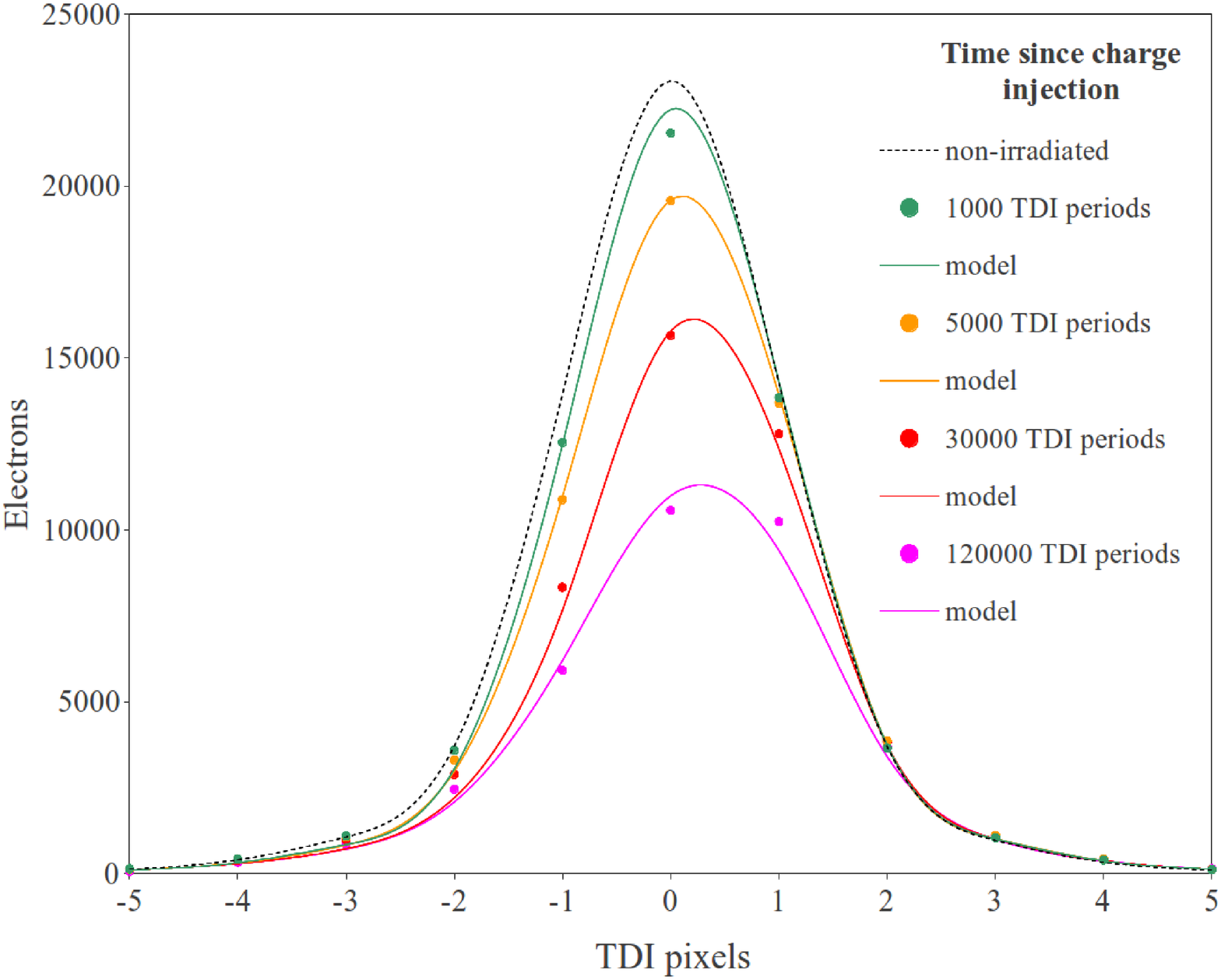}
\includegraphics[width=80mm]{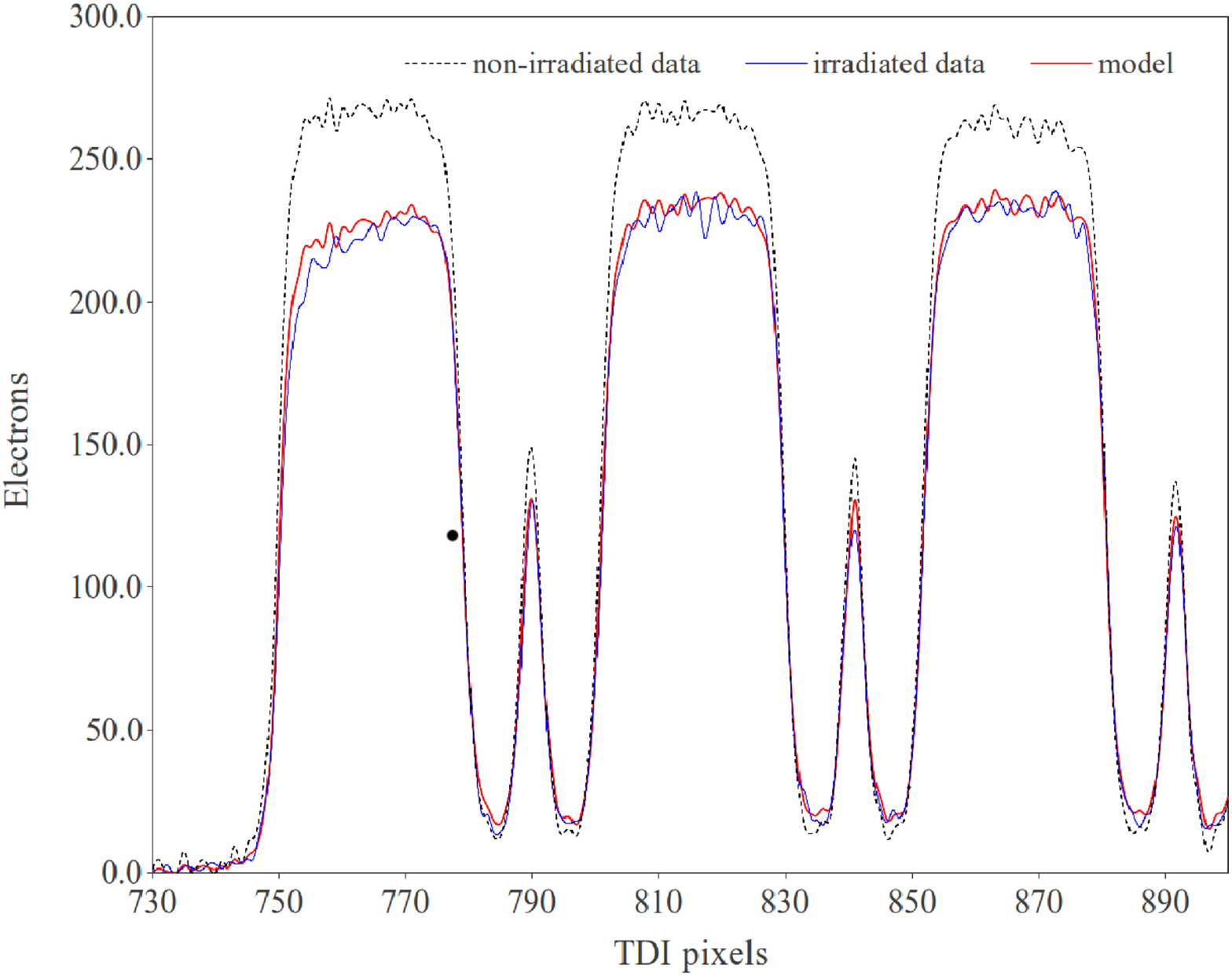}
\caption{{\bf{Left:}} Representive fits to {\Gaia}  test data for {\Gaia} magnitude $\sim$15 point sources with varying time since charge injection. The non-irradiated curve is derived from data acquired on a  section of a CCD which has not undergone radiation damage. The other curves are derived from data acquired from a radiation-damaged section of the same detector, with a uniform damage level throughout this section.  {\bf{Right:}}  Similarly, a section of {\Gaia} dispersed test data for {\Gaia} magnitude $\sim$10 and with a diffuse optical background of 0.77 \unit{e^{-} pix^{-1} s^{-1}} . \textcolor{black}{These data are representative of what will be observed with the Radial Velocity Spectrometer (RVS) instrument onboard {\Gaia}. For the fits shown in both plots model parameters are derived using a global fit to all test data from magnitude 13.25 to magnitude 20 and with delays since prior charge-injections from 30\,ms to 120\,s. Note that readout direction is to the left for both plots.}}
\label{fig:astro15}
\end{figure*}

It is beyond the scope of this work to provide a detailed analysis of the application of the model to {\Gaia} test data. Instead, we provide a  limited graphical indication that the model is qualitatively capable of reproducing trapping effects observed in test data. 

Figure~\ref{fig:astro15} (left panel) indicates the distortion observed in the parallel direction in {\Gaia} test data as a function of  time since charge injection. The datapoints are measured data and the solid lines are CDM model results. Note that a single set of trap parameter values was used to generate all of these model curves. Figure~\ref{fig:astro15} (right panel)  shows the fit of the model to radiation damaged dispersed spectral data (similar to what will be observed with the Radial Velocity Spectrometer (RVS) instrument onboard {\Gaia}).  Curves are shown for data taken on  non-irradiated and irradiated sections of a CCD,  with the output of the CDM also over-plotted.   Again, a single set of model parameters has been selected giving a reasonable fit to all data. The CDM has also been used to examine the effects of radiation damage on CTI in Euclid \cite[e.g., see][]{2012SPIEEuclidL,2013MNRASEuclidC} CCDs.

\section{Conclusions} \label{sect:conclusions}

The ESA astrometric mission {\Gaia}, scheduled for launch in 2013, will be subject  to a systematic bias in its measurements due to radiation damage-induced CTI in its CCD detectors. In order to sufficiently calibrate out this bias it  is planned to forward model the charge distortion to the CCD images at the CCD sample level. 
The  use of an analytical charge distortion model, which must be computationally inexpensive, is fundamental to {\Gaia} reaching its scientific  goals. It treats electron trapping and de-trapping within each CCD column (and serial register) in a single step ensuring that it is  computationally fast, although still based on physical theory. A description of the model is presented, and its ability to qualitatively reproduce the CTI effects in {\Gaia} test data is demonstrated. The model can be applied to CCDs operating in both TDI and imaging mode and hence is fully general.

\section*{Acknowledgments}
We would like to thank the {\Gaia}  Radiation Task Force of the {\Gaia} Data Processing and Analysis Consortium for guidance, advice and data-analysis expertise. We also wish to thank  EADS Astrium and ESA project for providing radiation test data and for responding readily to all of our requests for further information as well as the  anonymous referee for  insightful feedback.

\bibliographystyle{mn2e}
\bibliography{short}


\end{document}